\documentclass{ws-procs975x65}

\usepackage{amsmath}

\begin{document}

\title{Numerical wave optics and the lensing of gravitational waves by globular clusters}

\author{Andrew J. Moylan, David E. McClelland, Susan M. Scott and Antony C. Searle}

\address{Centre for Gravitational Physics, Department of Physics, The Australian National University, Canberra ACT 0200, Australia \\
\email{andrew.moylan@anu.edu.au, david.mcclelland@anu.edu.au, susan.scott@anu.edu.au, antony.searle@anu.edu.au}}

\author{G. V. Bicknell}

\address{Research School of Astronomy and Astrophysics, The Australian National University, Canberra ACT 2611, Australia \\
\email{geoff@mso.anu.edu.au}}

\begin{abstract}

We consider the possible effects of gravitational lensing by globular clusters on gravitational waves from asymmetric neutron stars in our galaxy. In the lensing of gravitational waves, the long wavelength, compared with the usual case of optical lensing, can lead to the geometrical optics approximation being invalid, in which case a wave optical solution is necessary. In general, wave optical solutions can only be obtained numerically. We describe a computational method that is particularly well suited to numerical wave optics. This method enables us to compare the properties of several lens models for globular clusters without ever calling upon the geometrical optics approximation, though that approximation would sometimes have been valid. Finally, we estimate the probability that lensing by a globular cluster will significantly affect the detection, by ground-based laser interferometer detectors such as LIGO, of gravitational waves from an asymmetric neutron star in our galaxy, finding that the probability is insignificantly small.

\end{abstract}

\section{Introduction}

In this article we consider wave optical effects in the gravitational lensing of gravitational waves by globular clusters, and determine the possible effect of such lensing on gravitational waves detected by Earth-based interferometers. Wave effects in gravitational lensing are also relevant in contexts other than the lensing of gravitational waves, such as the `femtolensing' of gamma-ray bursts by hypothetical compact objects of very low mass, and have been studied by various authors; see Nakamura and Deguchi\cite{nakamura_wave_1999} and references therein.

Several authors have considered the possible wave optical effects of lensing on gravitational waves.\cite{nakamura_gravitational_1998,ruffa_gravitational_1999,baraldo_gravitationally_1999,takahashi_wave_2003,takahashi_quasi-geometrical_2004,macquart_scattering_2004,takahashi_amplitude_2006}
Beginning with Nakamura\cite{nakamura_gravitational_1998}, many of these studies have employed the Kirchhoff-type diffraction integral found in the book by Schneider et al.\cite{schneider_gravitational_1992} and re-derived by Nakamura and Deguchi\cite{nakamura_wave_1999} using a path integral approach. Takahashi and Nakamura\cite{takahashi_wave_2003} investigated the wave optical properties of two simple lens profiles, the singular isothermal sphere and the point-mass, and used them to model dark matter halos and compact objects (such as black holes), respectively, obtaining an estimate of the effects of lensing by these objects on gravitational waves in the sensitive frequency range of the proposed space-based detector LISA.

In the geometrical optics limit, the lensing of gravitational waves by a variety of lens models has been considered by other authors including Arnaud-Varvella et al.\cite{arnaud-varvella_increase_2004}\ and Seto\cite{seto_strong_2004}. Arnaud-Varvella et al.\ found that the geometrical optics approximation is applicable in the frequency range to which Earth-based interferometers are sensitive ($10^1$-$10^4 \, \text{Hz}$) for lens masses $\gtrsim 10^6 \, M_\odot$. (For a slightly different criterion, $\text{lens mass} \gtrsim 10^8 \, M_\odot ({f}/{\text{mHz}})^{-1}$, see Suyama et al.\cite{suyama_wave_2005}\ and references therein.) The meaning of this criterion is unclear, however, for common lens models that do not have a finite total mass, such as isothermal spheres, and the geometrical optics approximation remains, nonetheless, invalid when the images lie near caustics of the lens. In deriving the geometrical optics limit from the diffraction integral, Nakamura and Deguchi have obtained a condition on the wave frequency and lensing configuration that determines when the geometrical optics approximation is valid in general (Equation (3.3) of reference \refcite{nakamura_wave_1999}).

The diffraction integral relevant to wave optical lensing must, in general, be computed numerically, and is rapidly oscillatory, making it difficult to evaluate. The field of highly oscillatory integration, which is currently very active,\cite{iserles_highly_2005} offers promising methods for these types of integrals. After reviewing wave optics in gravitational lensing in Section~\ref{S:wave optics}, we describe in Section~\ref{S:numerical wave optics} an algorithm,\cite{moylan_highly_2007} based on a method originally due to Levin\cite{levin_procedures_1982}, capable of very efficiently computing the diffraction integrals necessary in analysing the lens models considered in later sections. In Section~\ref{S:lens models} we describe a variety of possible lens profiles for modelling globular clusters, and explore their wave optical properties. In Section~\ref{S:detection} we use our results to estimate the probability that lensing by globular clusters in our galaxy can significantly affect the detection by Earth-based interferometers of gravitational waves from asymmetric neutron stars in our galaxy.

We use geometric units in which $c = G = 1$. Our notation, which closely follows that of Takahashi,\cite{takahashi_wave_2004} is summarised in Table~\ref{T:notation}.

\begin{table}
  \tbl{Summary of notation.}
  {
    \begin{tabular}{| r | l |}
      \hline
      $D_L$ & (angular diameter) distance from observer to lens \\
      $D_{LS}$ & distance from lens to source \\
      $D_L + D_{LS} \equiv D_S$ & distance from observer to source \\
      $M_L$ & lens mass \\
      \hline
      $\mathbf {\hat x}$ & position in the lens plane \\
      $\mathbf {\hat y}$ & displacement of source from optical axis \\
      $\hat w$ & angular frequency of wave \\
      $\hat \psi (\mathbf {\hat x})$ & 2-dimensional lensing potential \\
      \hline
      $\xi_0$ & length normalisation constant \\
      $\mathbf {\hat x} / \xi_0 \equiv \mathbf x$ & dimensionless position in lens plane \\
      $(D_L / D_S \xi_0) \mathbf {\hat y} \equiv \mathbf y$ & dimensionless displacement of source from optical axis \\
      $(D_S / D_L D_{LS}) \xi_0^2 \hat w \equiv w$ & dimensionless angular frequency \\
      $(D_L D_{LS} / D_S \xi_0^2) \hat \psi (\mathbf {\hat x}) \equiv \psi (\mathbf x)$ & dimensionless 2-dimensional lensing potential \\
      \hline
    \end{tabular}
  }
  \label{T:notation}
\end{table}

\section{Wave optics in gravitational lensing}
\label{S:wave optics}

In this section we briefly review the essential parts of the theory of wave optics in gravitational lensing, and present the Kirchhoff-type diffraction integral, Equation~\eqref{E:kirchhoff-type diffraction integral} below, that is the starting point for our investigations of the wave optical properties of lens models discussed in subsequent sections. For a more detailed derivation of the diffraction integral, see e.g. Schneider et al.\cite{schneider_gravitational_1992}, Nakamura and Deguchi\cite{nakamura_wave_1999} and Takahashi\cite{takahashi_wave_2004}.

Assuming that the polarisation of the incoming wave is not significantly altered by its interaction with the lens, the amplitude $\phi$ of the wave satisfies the frequency domain scalar wave equation
\begin{equation}
  \label{E:scalar wave equation}
  \left( \nabla^2 + {\hat w}^2 \right) \phi = 4 {\hat w}^2 U \phi,
\end{equation}
where $\hat w$ is the angular frequency of the wave and $U \ll 1$ is the weak gravitational potential of the lens:
\begin{equation}
  \label{E:poissons-equation}
  \nabla^2 U = 4 \pi \rho,
\end{equation}
where $\rho$ is the mass density of the lens.

Assuming further that the thin lens approximation applies, the lens is totally characterised by its 2-dimensional lensing deflection potential
\begin{equation}
  \label{E:deflection-potential}
  \hat{\psi} (\mathbf {\hat x}) = 2 \int_{-\infty}^\infty U(\mathbf {\hat x}, z) \, dz,
\end{equation}
where the integral is over the optical axis, and $\mathbf{\hat x}$ is the position on the lens plane. (Suyama et al.\cite{suyama_wave_2005} have tested the validity of this thin lens approximation in wave optical gravitational lensing, finding that it is valid for a variety of lens models under reasonable astrophysical parameters.)

From these assumptions follows the Kirchhoff-type diffraction integral over the lens plane,
\begin{equation}
  \label{E:kirchhoff-type diffraction integral}
  F (w, \mathbf y) = \frac{w}{2 \pi i} \int_{{\mathbb R}^2} \exp \left[ i w T (\mathbf x, \mathbf y) \right] \, d^2 \negthinspace x,
\end{equation}
giving the amplification factor $F$, which is the ratio of the amplitude $\phi$ at the observer to the amplitude $\phi_0$ at the observer in the absence of a lens. In Equation~\eqref{E:kirchhoff-type diffraction integral} we have switched to dimensionless variables scaled by a length normalisation constant $\xi_0$ (see Table~\ref{T:notation}), specified separately for each lens model studied below. (In general the dimensionless quantities of Table~\ref{T:notation} include a dependence on the lens redshift $z_L$, but in this article we do not consider lenses at cosmological distances.)
\begin{equation}
  \label{E:time-delay}
  T (\mathbf x, \mathbf y) = \frac{1}{2} {\left| \mathbf x - \mathbf y \right|}^2 - \psi (\mathbf x)
\end{equation}
is the dimensionless optical time delay along a path from the source at $\mathbf y$ to the observer via a point $\mathbf x$ on the lens plane. The first term in Equation~\eqref{E:time-delay} is the geometric time delay in the absence of a lens, and the second term is the time delay due to the gravitational potential of the lens.

When the lens model is axisymmetric ($\psi (\mathbf x) = \psi (x)$ where $x = | \mathbf x |$), which is the case for all of the lens models considered in this article, Equation~\eqref{E:kirchhoff-type diffraction integral} can be rewritten as a 1-dimensional integral involving a Bessel function:
\begin{equation}
  \label{E:axisymmetric-diffraction-integral}
  F (w, y) = -i w \exp (i w y^2 / 2) \int_0^\infty x J_0 (w x y) \exp \left[i w \left(\frac 1 2 x^2 - \psi (x) \right) \right] dx.
\end{equation}

\subsection{Comment on convergence of the diffraction integral}

The 2-dimensional diffraction integral of Equation~\eqref{E:kirchhoff-type diffraction integral} may be described as an `integral over the lens plane' and written as $\int_{{\mathbb R}^2}$. The integral is not, however, absolutely convergent, and therefore the particular order in which the integration over ${\mathbb R}^2$ is performed determines the value to which it converges. For example, consider evaluating the integral in polar coordinates $(r, \theta)$ by integrating over $r$ first and over $\theta$ second:
\begin{equation}
  \label{E:invalid-order-of-integration}
  \int_{\mathbb{R}^2} \cdots d^2 \negthinspace x \equiv \int_0^{2 \pi} \int_0^\infty \cdots r \, dr \, d\theta = \int_0^{2 \pi} \left[ \lim_{R \to \infty} \int_0^R \cdots r \, dr \right] d\theta.
\end{equation}
(Note that it is necessary to make explicit that $\int_0^\infty dr = \lim_{R \to \infty} \int_0^R dr$, because the integral over $r$ is itself not absolutely convergent.) The integral of Equation~\eqref{E:invalid-order-of-integration} does not converge, because the limit in brackets does not exist; rather, at a fixed value of $\theta$, the partial integral $\int_0^R dr$ as a function of $R$ typically oscillates indefinitely between two fixed values.

When the integral is evaluated in polar coordinates by integrating over $\theta$ first, however,
\begin{equation}
  \label{E:valid-order-of-integration}
  \int_{\mathbb{R}^2} \cdots d^2 \negthinspace x \equiv \int_0^\infty \int_0^{2 \pi} \cdots r \, d\theta \, dr = \lim_{R \to \infty} \int_0^R \left[ \int_0^{2 \pi} \cdots d\theta \right] \, r \, dr,
\end{equation}
the limit as $R \to \infty$ generally does exist, and this definition of the diffraction integral \emph{does} approximately solve the scalar wave equation, Equation~\eqref{E:scalar wave equation}. Performing the integral as in Equation~\eqref{E:valid-order-of-integration} is implicit in the derivation of Equation~\eqref{E:axisymmetric-diffraction-integral}, the 1-dimensional form for axisymmetric lenses. There, $\int_0^\infty dx$ means $\lim_{X \to \infty} \int_0^X dx$ as usual, and the limit generally exists.

\section{Numerical wave optics}
\label{S:numerical wave optics}

The diffraction integral, in the form of Equation~\eqref{E:kirchhoff-type diffraction integral} or Equation~\eqref{E:axisymmetric-diffraction-integral}, contains an infinite number of oscillations in the (infinite) range of integration. More importantly, it is \emph{rapidly} oscillatory in the sense that the region of integration that contributes significantly to the integral may contain many oscillations. It is therefore computationally expensive to evaluate using traditional interpolatory methods such as Gaussian quadrature.

In the context of femtolensing of gamma-ray bursts, Ulmer and Goodman\cite{ulmer_femtolensing_1995} (see also Nakamura and Deguchi\cite{nakamura_wave_1999}) developed a method for evaluating Equation~\eqref{E:kirchhoff-type diffraction integral} based on first integrating over contours $T (\mathbf x, \mathbf y) = \tau$ of the time delay, and then integrating over $\tau$. The contours of $T (\mathbf x, \mathbf y)$ must in general be determined numerically, usually by stepping outward from the geometrical optics images, which must therefore be located in advance.

Takahashi\cite{takahashi_wave_2004} used an asymptotic method for solving Equation~\eqref{E:axisymmetric-diffraction-integral} in which, after a change of variable $z = x^2 / 2$, the integral is split into $\int_0^\infty dz = \int_0^b dz + \int_b^\infty dz$. The first integral is evaluated using standard interpolatory quadrature, while the second integral is repeatedly integrated by parts, leading to an asymptotic series that converges for large enough $b$. A disadvantage of this method is that the first integral, though over a finite range, may nonetheless contain very many oscillations of the integrand, making the interpolatory quadrature computationally expensive.

\subsection{Highly oscillatory integration}

The field of highly oscillatory integration is concerned with the efficient evaluation of integrals like Equations~\eqref{E:kirchhoff-type diffraction integral} and~\eqref{E:axisymmetric-diffraction-integral}, along with related problems such as the efficient solution of highly oscillatory ordinary differential equations. Numerous techniques have been developed; see the review by Iserles et al.\cite{iserles_highly_2005} The \emph{Levin-type methods}\cite{levin_procedures_1982,levin_fast_1996,levin_analysis_1997}, most recently studied by Olver,\cite{olver_moment-free_2006,olver_quadrature_2006} are particularly well suited to numerical wave optics in gravitational lensing. Our algorithm is based on these methods.

Levin's original method\cite{levin_procedures_1982} addresses integrals of the form
\begin{equation}
  \label{E:oscillatory-exp-integral}
  \int_a^b f(x) \exp (i g(x)) \, dx,
\end{equation}
where $f$ and $g$ are not rapidly oscillatory. Levin observed that, among the antiderivatives of $f(x) \exp (i g(x))$, which are themselves rapidly oscillatory, there is one antiderivative that can be written as $F(x) \exp (i g(x))$ where $F (x)$ is not rapidly oscillatory. $F (x)$ satisfies the differential equation
\begin{equation}
  \label{E:levin-equation}
  F'(x) + i g'(x) F(x) = f(x),
\end{equation}
and may be approximated by the unique order-$n$ \emph{collocation} polynomial that satisfies Equation~\eqref{E:levin-equation} at $n$ arbitrarily chosen points within the region of integration.

Importantly, for our purposes, Levin subsequently showed\cite{levin_fast_1996} that the oscillator, $\exp (i \, \cdot)$, of Equation~\eqref{E:oscillatory-exp-integral} may, in general, be replaced with any function satisfying the criterion that it can be written as an element of a vector $\boldsymbol \omega (x)$ of oscillators satisfying the linear differential relation
\begin{equation}
  \label{E:general-oscillator}
  \boldsymbol \omega' (x) = \mathbf A (x) \boldsymbol \omega (x),
\end{equation}
where $\mathbf A (x)$ is a matrix of non-rapidly-oscillatory functions. (This criterion is equivalent to the oscillator satisfying a homogeneous linear differential equation of some order; see Chung et al.\cite{chung_method_2000}, who also show that the requirement of homogeneity may, in general, be dropped.) The product of any two oscillators satisfying this criterion also satisfies the criterion, and the composition of an oscillator satisfying the criterion with a non-rapidly oscillatory function also satisfies the criterion (Levin\cite{levin_fast_1996} Lemmas~1 and~2). The Bessel function $J_0$ satisfies the criterion, with Equation~\eqref{E:general-oscillator} becoming
\begin{equation}
  \left( \begin{array}{c} J_0' (x) \\ J_1' (x) \end{array} \right)
  =
  \left( \begin{array}{ccc} 0 & & -1 \\ 1 & & -1/x \end{array} \right)
  \left( \begin{array}{c} J_0 (x) \\ J_1 (x) \end{array} \right)
  .
\end{equation}
The oscillator of Equation~\eqref{E:axisymmetric-diffraction-integral} therefore also satisfies the criterion, and the generalised form of Levin's method may be applied to that integral.

\subsection{Automatic integration algorithm based on Levin's method}

Levin showed that his method was able to accurately solve oscillatory integrals with many fewer function evaluations than traditional numerical quadrature schemes. We have developed an algorithm,\cite{moylan_highly_2007} in the form of a Mathematica package \texttt{LevinIntegrate}, that augments Levin's method with two crucial components of modern automatic integrators:

\begin{enumerate}

\item \textbf{Adaptive recursive subdivision} of the range of integration. This is both (i) necessary, because the $n$ point collocation solution to Equation~\eqref{E:levin-equation} involves inverting an $n \times n$ matrix, which becomes prohibitively computationally expensive as $n$ is increased in search of higher accuracy, and (ii) desirable, because the function $F (x)$ may be well approximated by piecewise low-order polynomials, but may not be well approximated by a single higher order polynomial. `Adaptive' refers to the algorithm selectively subdividing those regions of integration that are estimated to contribute most to the total remaining error.

\item \textbf{Automatic change of integration variable} to compactify any infinite range of integration, and to remove singularities of the function $f (x)$ at the endpoints of the range of integration.

\end{enumerate}

The \texttt{LevinIntegrate} package automatically applies Lemmas~1 and~2 of Levin\cite{levin_fast_1996} to determine the relevant vector $\boldsymbol \omega (x)$ of oscillators and corresponding matrix $\mathbf A (x)$ of Equation~\eqref{E:general-oscillator}, and tries to estimate the value of the given integral to within any requested precision.

\begin{figure}
  \begin{center}
    \psfig{file=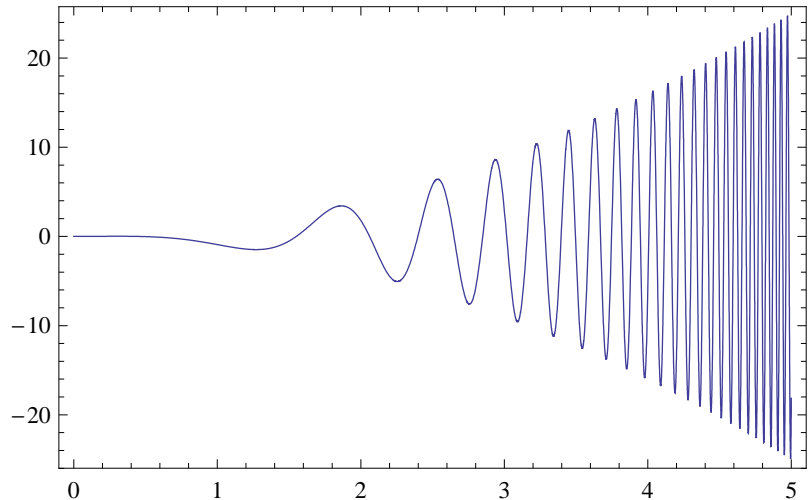}
    \caption{The real part of the highly oscillatory function $x^2 \exp (i \exp x)$.}
    \label{F:pathological-integrand}
  \end{center}
\end{figure}

By way of example, consider the apparently pathological test integral
\begin{equation}
  \label{E:pathological-integral}
  I(b) = \int_0^b h(x) \, dx, \quad h(x) = x^2 \exp (i \exp (x)),
\end{equation}
which oscillates with an exponentially increasing frequency. The real part of $h (x)$ is shown in Figure~\ref{F:pathological-integrand}.

Mathematica's highly optimised \texttt{NIntegrate} routine, which uses Gaussian (interpolatory) quadrature, is superior to \texttt{LevinIntegrate} for non-rapidly-oscillatory integrals. Running on a current CPU, \texttt{NIntegrate} evaluated $I (5)$ to 6 significant figures in $0.04\,\text{s}$, evaluating $h(x)$ 1628 times. \texttt{LevinIntegrate} evaluated $I(5)$ to the same precision in $0.25\,\text s$, evaluating $h(x)$ 353 times. The execution time per evaluation of $h(x)$ is expected to be higher for $\texttt{LevinIntegrate}$ than for $\texttt{NIntegrate}$ because of the comparatively costly matrix inversions performed by the former, and because of the superior optimisation built into the latter.

For rapidly oscillatory integrals, \texttt{LevinIntegrate} is superior to \texttt{NIntegrate}. \texttt{NIntegrate} evaluated $I (12)$ to 6 significant figures in $58\,\text s$, evaluating $h(x)$ 1925583 times. \texttt{LevinIntegrate} evaluated $I(12)$ to the same precision in $0.25\,\text s$, evaluating $h(x)$ 353 times (the same amount of computation as for $I(5)$). Evaluating $I(b)$ for different values of $b$ will, of course, not always require exactly 353 evaluations of $h(x)$. It is, however, a feature of Levin-type methods that the computational cost is approximately independent of (rather than proportional to) the frequency of the integrand.

Levin's original method explicitly excludes cases in which the `phase' function ($g(x)$ in Equation~\eqref{E:oscillatory-exp-integral}) has stationary points within the region of integration. Stationary points correspond to `images' in the geometrical optics approximation, and the contribution to the integral is largest for regions around these points. Nonetheless, the \texttt{LevinIntegrate} algorithm handles stationary points. This is because (i)~the adaptive recursive subdivision employed in \texttt{LevinIntegrate} effectively automatically locates stationary points in the phase function; and (ii)~Levin's collocation method is, in fact, competitive with traditional interpolatory integration methods in small regions containing these points. See Moylan et al.\cite{moylan_highly_2007} for details.

\section{Wave optical properties of various lens models for globular clusters}
\label{S:lens models}

Using \texttt{LevinIntegrate} to evaluate Equation~\eqref{E:axisymmetric-diffraction-integral}, we have numerically investigated a variety of possible simple lens models for globular clusters. In this section we describe the models and compare their wave-optical properties. The lens models we consider are summarised in Table~\ref{T:lens-model-properties}.

\begin{table}
  \tbl{Properties of simple lens models for globular clusters.}
  {
    \begin{tabular}{| l | c | l | l |}
      \hline
      model & parameters & total mass & central density \\
      \hline
      point-mass & 1 & finite & infinite \\
      Plummer & 2 & finite & finite \\
      singular isothermal sphere & 1 & infinite & infinite \\
      non-singular isothermal sphere & 2 & infinite & finite \\
      \hline
    \end{tabular}
  }
  \label{T:lens-model-properties}
\end{table}

When we apply the models to the lensing by specific globular clusters of gravitational waves from sources (such as asymmetric neutron stars) in our galaxy, unless otherwise stated we consider monochromatic gravitational waves at a typical frequency of $200\,\text{Hz}$ ($\omega_\text{GW} = (2 \pi) (200)\,\text{rad}/\text{s}$), which is in the most sensitive range for LIGO. We take globular cluster data from the catalogues published by Harris\cite{harris_catalog_2003} and Pryor et al.\cite{pryor_velocity_1993}, and from other sources as cited. Approximate parameters for the globular cluster M22, which lies in the galactic plane and is projected in front of the galactic bulge, are summarised in Table~\ref{T:M22-parameters}.\footnote{In Table~\ref{T:M22-parameters}, the core radius is the radius at which the observed surface brightness is half the central value, and the radial velocity dispersion is the standard deviation in the radial component of the velocity of the objects comprising M22.}

\begin{table}
  \tbl{Parameters of globular cluster M22.}
  {
    \begin{tabular}{| r | l |}
      \hline
      mass & $3 \times 10^5\,\text{$M_\odot$} = 4.4 \times 10^8\,\text{m}$ \\
      distance from Earth & $3.2\,\text{kpc}$ \\
      half-mass radius & $3\,\text{pc}$ \\
      core radius & $1.3\,\text{pc}$ \\
      observed radial velocity dispersion\cite{peterson_proper_1994} & $2.2 \times 10^{-5}$ \\
      \hline
    \end{tabular}
  }
  \label{T:M22-parameters}
\end{table}

\subsection{Point-mass lens}

The density profile of the point-mass lens, the simplest model for an object of total mass $M_L$, is
\begin{equation}
  \rho (\mathbf r) = M_L \delta (\mathbf r).
\end{equation}
We follow Takahashi et al.\cite{takahashi_wave_2003} in adopting the Einstein radius $\xi_E \equiv \sqrt {4 M_L D_L D_{LS} / D_S}$ as the length normalisation constant: $\xi_0 = \xi_E$. Then the lensing potential is $\psi (x) = \ln x$, and Equation~\eqref{E:axisymmetric-diffraction-integral} has the closed-form solution\cite{peters_index_1974,takahashi_wave_2003}
\begin{equation}
  F (w, y) =
\exp{\left[ \frac {\pi w} 4  + \frac {iw} 2 \left(\ln \frac w 2 - 2 \phi_m (y)\right) \right]} \Gamma (1 - \frac {iw} 2) {}_1 F_1 (\frac {iw} 2, 1, y^2 \frac {iw} 2),
\end{equation}
where $\phi_m (y) = (x_m - y)^2 / 2 - \ln x_m$, $x_m = (y + \sqrt{y^2 + 4}) / 2$, $\Gamma$ is the Euler gamma function, and ${}_1 F_1$ is the Kummer confluent hypergeometric function.

\begin{figure}
  \begin{center}
    \psfig{file=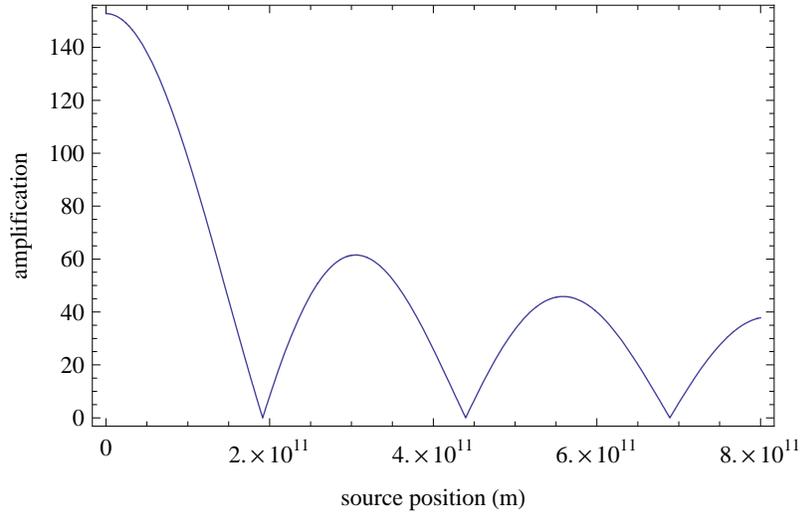}
    \caption{Amplification for waves from a source twice as distant as the globular cluster M22 lensed by a point-mass with the same mass and position as M22, as a function of the distance of the source from the optical axis.}
    \label{F:point-mass-M22-diffraction-pattern}
  \end{center}
\end{figure}

Figure~\ref{F:point-mass-M22-diffraction-pattern} shows part of the amplification diffraction pattern for lensing by the globular cluster M22 modelled as a point-mass, for a source twice as distant as M22.
Although the amplification is as large as 150, the point-mass lens is not expected to be a reasonable model for extended mass distributions like globular clusters, as we will see in the following sections.

\subsection{Plummer model}

The Plummer model\cite{plummer_problem_1911,binney_galactic_1987} may be defined in terms of the Plummer radius $a$ and the total mass $M_L$. The density is
\begin{equation}
  \rho (\mathbf r) = \frac {3 M_L} {4 \pi a^2} \left(1 + \frac {r^2} {a^2 } \right)^{-5/2}.
\end{equation}
Like the point-mass lens, and unlike the singular and non-singular isothermal spheres, the Plummer model has a finite total mass $M_L$. A fraction $\sqrt 2 / 4 \simeq 35 \%$ of the total mass is contained within $a$, and the half-mass radius is
\begin{equation}
  \label{E:plummer-half-mass-radius}
  r_\text{half} = \frac{1 + 2^{1/3}}{\sqrt 3} a \simeq 1.3 \, a.
\end{equation}

One natural choice for the length normalisation constant is $\xi_0 = a$ (Schneider et al.\cite{schneider_gravitational_1992}\ p.\,245), but here we choose $\xi_0 = \xi_E = \sqrt {4 M_L D_L D_{LS} / D_S}$, the same as for a point-mass lens of the same mass, for ease of comparison between the Plummer and point-mass lens models. In terms of the dimensionless central surface mass density parameter
\begin{equation}
  \kappa_0 \equiv \frac {4 D_L D_{LS}} {D_S} \frac {M_L} {a^2} = \left( \frac{\xi_E}{a} \right)^2,
\end{equation}
the lensing deflection potential for the Plummer model is
\begin{equation}
  \psi (x) = \frac 1 2 \ln (1 + \kappa_0 x^2).
\end{equation}
When $\kappa_0 > \sim 1.7$, most of the mass of the Plummer profile is contained within the Einstein radius $\xi_E$ corresponding to a point-mass model of mass $M_L$. In this case, we expect the Plummer model to have similar lensing properties to the corresponding point-mass. Conversely, when $\kappa_0 \leq 1$, we expect the lensing properties to differ from those of the corresponding point-mass. For this lens model, and for all the subsequent lens models we consider, Equation~\eqref{E:axisymmetric-diffraction-integral} (which gives the amplification factor) must be integrated numerically.

\begin{figure}
  \begin{center}
    \psfig{file=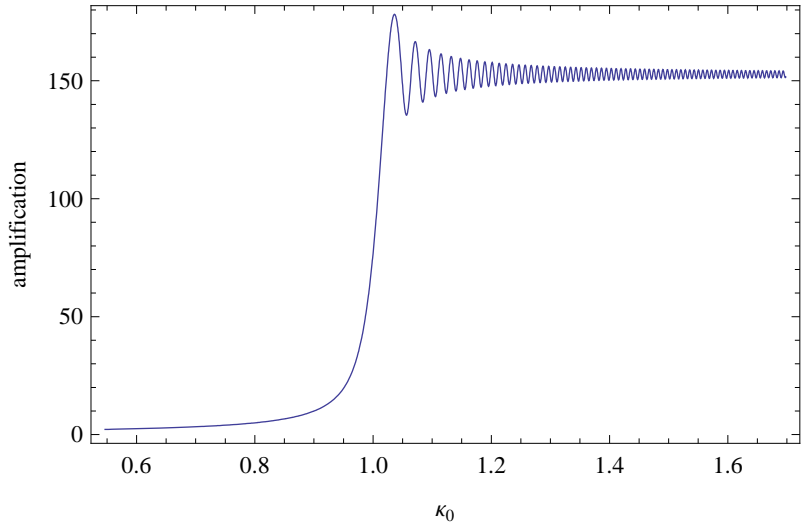}
    \caption{Amplification for waves from a source lying on the optical axis, lensed by the globular cluster M22 modelled as a Plummer profile, as a function of the dimensionless central surface mass density parameter $\kappa_0$.}
    \label{F:plummer-v-point-mass-on-axis}
  \end{center}
\end{figure}

Figure~\ref{F:plummer-v-point-mass-on-axis} shows, as a function of the parameter $\kappa_0$, the height of the central maximum of the diffraction pattern, for the same lensing configuration as Figure~\ref{F:point-mass-M22-diffraction-pattern}, when M22 is modelled using a Plummer profile instead of a point-mass.
Note that, for some values of the core radius satisfying $a \lesssim \xi_E$ (that is, $\kappa_0 \gtrsim 1$), the on-axis amplification is somewhat \emph{larger} for a Plummer profile than for a point-mass of the same total mass.

From Equation~\eqref{E:plummer-half-mass-radius} and the half-mass radius for M22 (Table~\ref{T:M22-parameters}), we find that $a \simeq 7 \times 10^{16}\,\text{m}$ is an appropriate choice when modelling M22 as a Plummer profile. For lensing of waves from a source twice as distant as M22, this corresponds to $\kappa_0 = 1.7 \times 10^{-5}$. Comparison with Figure~\ref{F:plummer-v-point-mass-on-axis}, which spans the range $0.55 \leq \kappa_0 \leq 1.7$, shows that (if M22 is well-modelled by a Plummer profile) M22 is not well-modelled by a point-mass. In fact, for M22 modelled as a Plummer profile, the maximum amplification differs from unity by less than $10^{-4}$. Therefore, if M22 is well-modelled by a Plummer profile, lensing of gravitational waves by M22 cannot yield significant amplification.

\subsection{Singular isothermal sphere}
\label{S:sis}

Takahashi et al.\cite{takahashi_wave_2003} used the singular isothermal sphere (SIS) profile to model dark matter halos, and suggested it as a model for star clusters. For this model,
\begin{equation}
  \rho (\mathbf r) = \frac {v^2} {2 \pi r^2},
\end{equation}
where $v$ is the (dimensionless) line-of-sight velocity dispersion.\cite{binney_galactic_1987} With the length normalisation constant chosen as $\xi_0 = 4 \pi v^2 D_L D_{LS} / D_S$, the lensing potential is $\psi (x) = x$.

\begin{figure}
  \begin{center}
    \psfig{file=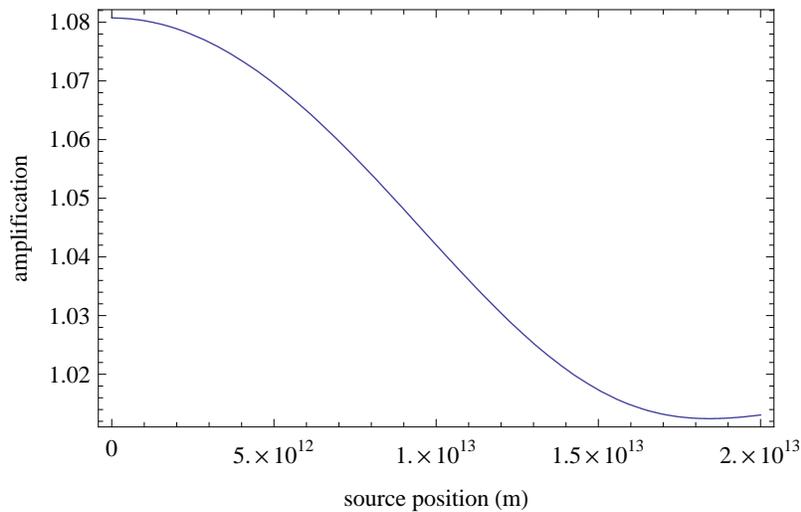}
    \caption{Amplification diffraction pattern for the same lensing configuration as Figure~\ref{F:point-mass-M22-diffraction-pattern}, but with the globular cluster M22 modelled as a singular isothermal sphere (SIS).}
    \label{F:sis-M22-diffraction-pattern}
  \end{center}
\end{figure}

Figure~\ref{F:sis-M22-diffraction-pattern} shows the amplification diffraction pattern for the same lensing configuration as Figure~\ref{F:point-mass-M22-diffraction-pattern}, except with M22 modelled as a SIS profile instead of a point-mass profile. The maximum (on-axis) amplification is less than 1.1. Even for waves at a higher frequency of 1000\,Hz (the upper limit of LIGO's sensitive range), the maximum (i.e., on-axis) amplification for M22 modelled as a SIS profile with a velocity dispersion of $2.2 \times 10^{-5}$ is less than 1.3.

For globular clusters with higher velocity dispersions, such as the largest Milky Way globular cluster, $\omega$-Centauri ($D_L = 5.6\,\text{kpc}$, $v = 5.6 \times 10^{-5}$), the corresponding SIS profile can have a higher maximum amplification: above 2 (for $D_{LS} \gtrsim D_L$), increasing to above 6 for waves of frequency 1000\,Hz. $\omega$-Centauri lies far from the galactic disk, however, and indeed we find no globular clusters lying nearby in the galactic plane with an observed velocity dispersion greater than that of M22.

\subsection{Non-singular isothermal sphere}

The non-singular isothermal sphere (NSIS) may be parameterised by its finite central density $\rho_0$ and a characteristic scale radius (`King radius') $r_0$.\cite{binney_galactic_1987} The density profile $\rho (\mathbf r)$ is the solution to the differential equation
\begin{equation} \label{E:nsis-definition}
  \frac{d}{d \tilde r} \left[ {\tilde r}^2 \frac{d \ln \tilde \rho}{d \tilde r} \right] = -9 {\tilde r}^2 \tilde \rho,
  \quad
  \tilde \rho (0) = 1,
  \quad
  \tilde \rho ' (0) = 0,
\end{equation}
where
\begin{equation}
  \tilde \rho \equiv \rho / \rho_0, \quad \tilde r \equiv r / r_0.
\end{equation}

Equation~\eqref{E:nsis-definition} has no closed form solution; it must be integrated numerically, to obtain (typically) a piecewise polynomial approximation of the function $\tilde \rho (\tilde r)$. Table~4-1 of Binney and Tremaine\cite{binney_galactic_1987} gives approximate values for $\log_{10} {\tilde \rho}$ and $\log_{10} (\Sigma / r_0 \rho_0)$ as a function of $\tilde r$, where
\begin{equation}
\Sigma (r) = \int_{-\infty}^{\infty} \rho (\sqrt{r^2 + z^2}) \, dz
\end{equation}
is the surface mass density corresponding to $\rho$. Their computed values are not accurate in all of the decimal places to which they are given. We have computed a more accurate version of Table~4-1 of Binney and Tremaine, which appears as Table~\ref{T:nsis-values} in Appendix~\ref{A:nsis-table}.

In order to study the lensing properties of the NSIS profile, the lensing deflection potential $\psi(x)$ must be found numerically. For a general numerically defined density profile, this may be accomplished by (i)~numerically solving the differential equation of Equation~\eqref{E:poissons-equation}, and then (ii)~numerically integrating Equation~\eqref{E:deflection-potential}. For the NSIS profile, however, a simple closed-form relation between $U$ and $\rho$ exists,\cite{binney_galactic_1987} so the first step may be skipped for that profile.

\begin{figure}
  \begin{center}
    \psfig{file=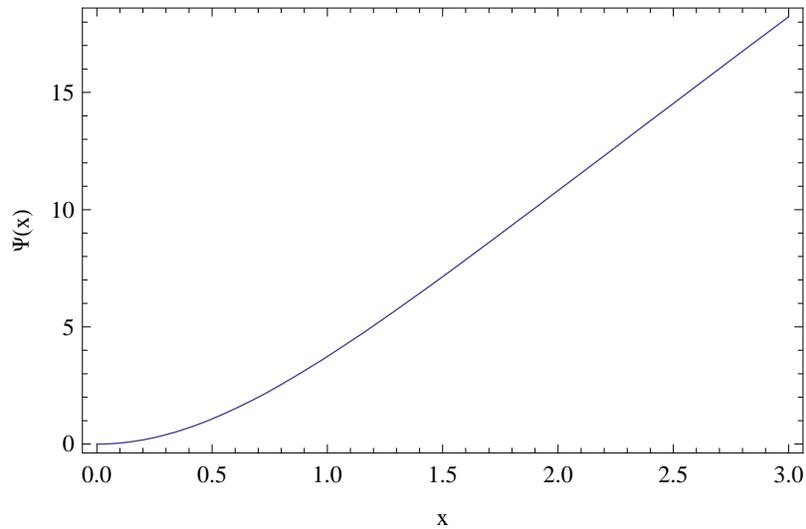}
    \caption{The numerically determined $\Psi$ of Equation~\eqref{E:nsis-defletion-potential}, to which the dimensionless lensing potential of the non-singular isothermal sphere (NSIS) is proportional, as a function of dimensionless radius $x$ in the lens plane. The asymptotic slope is $2 \pi$.}
    \label{F:nsis-deflection-potential}
  \end{center}
\end{figure}

That $\psi(x)$ is only defined numerically presents no particular impediment; Equation~\eqref{E:axisymmetric-diffraction-integral} may still be solved numerically via \texttt{LevinIntegrate}, just as for lens models for which $\psi(x)$ has a closed-form solution. We use the characteristic radius $r_0$ as the unit of distance normalisation in the lens plane, and write the deflection potential as
\begin{equation}
  \label{E:nsis-defletion-potential}
  \psi(x) = \psi_0 \Psi (x),
\end{equation}
where the `characteristic (deflection) potential' is
\begin{equation}
  \label{E:characteristic-potential-defined}
  \psi_0 \equiv \frac{8 \pi}{9} \frac{D_L D_{LS}}{D_S} \rho_0^2 r_0^3,
\end{equation}
and the numerically determined function $\Psi(x)$ is plotted in Figure~\ref{F:nsis-deflection-potential}. The velocity dispersion $v$ of a NSIS profile satisfies\cite{binney_galactic_1987}
\begin{equation}
  \label{E:nsis-vel-dispersion-squared}
  v^2 = \frac {4 \pi}{9} \rho_0 r_0^2,
\end{equation}
which leads to the following equivalent definition for the characteristic potential in terms of only $v$ and $r_0$:
\begin{equation}
  \psi_0 \equiv \frac{9}{2 \pi} \frac{D_L D_{LS}}{D_S} \frac{v^4}{r_0}.
\end{equation}

\begin{figure}
  \begin{center}
    \includegraphics[width=12cm]{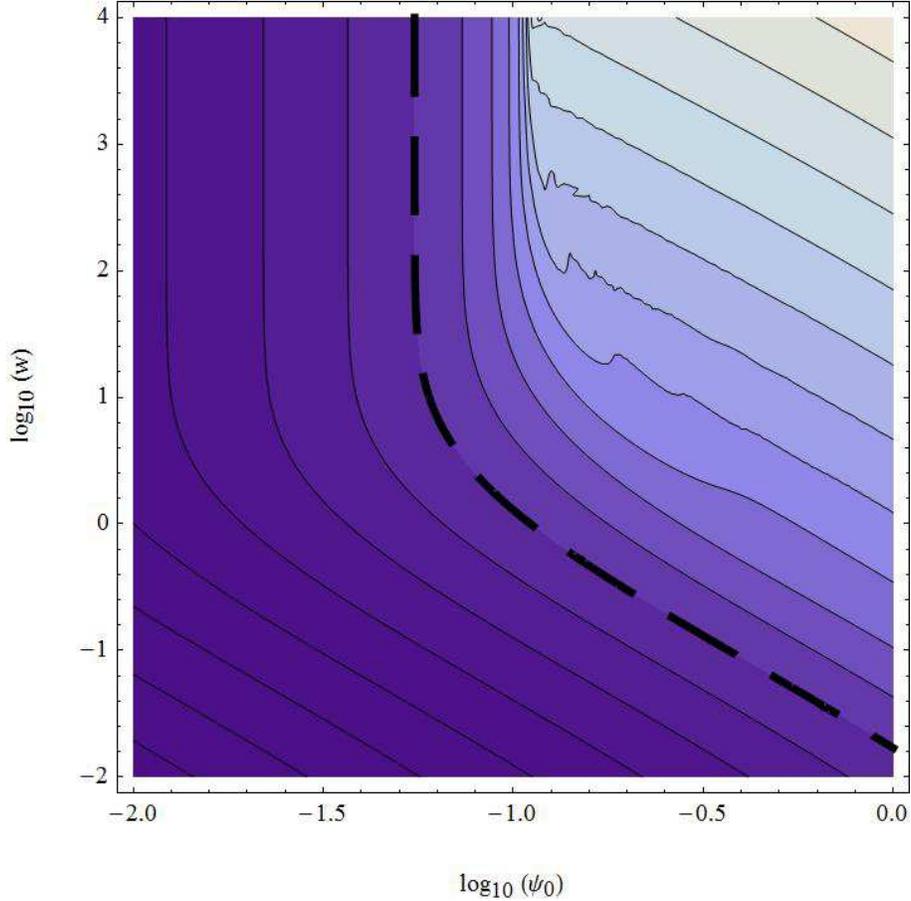}
    \caption{Contours of amplification for the non-singular isothermal sphere (NSIS) lens on-axis, as a function of the logarithm of the characteristic potential $\psi_0$ and the logarithm of the dimensionless angular frequency $w$. The contours are at values $1 + 2^n$ for integer values of $n$; the dashed contour is at the value 2 ($n = 0$). The contours with lighter shading are at higher values. For $\psi_0 > {\sim 0.11}$ ($\log_{10} \psi_0 > {\sim -0.96}$), the amplification diverges with increasing frequency (there is a caustic on-axis), and there are multiple images in the geometrical optics approximation.}
    \label{F:nsis-on-axis-contour}
  \end{center}
\end{figure}

Figure~\ref{F:nsis-on-axis-contour} shows the on-axis amplification for the NSIS lens model as a function of the parameters $w$ and $\psi_0$. For $\psi_0 < {\sim 0.11}$, the amplification remains finite as $w \to \infty$, and there is only a single image in the geometrical optics approximation. For $\psi_0 > {\sim 0.11}$, there are multiple images in the geometrical optics approximation, and the on-axis amplification diverges as $w \to \infty$, corresponding to the point $y = 0$ in the source plane being a caustic.

The core radius (at which the surface luminosity ($\propto$ surface mass density) is half the central value) of a NSIS profile is $r_\text{core} = 1.00344 r_0$. From this relation and Equation~\eqref{E:nsis-vel-dispersion-squared} we can determine suitable values for the parameters $\rho_0$ and $r_0$ to model a given globular cluster, by matching to the observed core radius and velocity dispersion. For M22, this implies $r_0 \simeq 4.1 \times 10^{16}\,\text{m}$ and $\rho_0 \simeq 2.1 \times 10^{-43}\,\text{m}^{-2}$. This central density parameter for M22 modelled as a NSIS profile is comparable to the central density for M22 modelled as a Plummer profile, $\rho_{\text{Plummer,M22}} (0) \simeq 2.9 \times 10^{-43}\,\text{m}^{-2}$, so we may expect insignificant amplification just as for the Plummer lens.

\begin{figure}
  \begin{center}
    \psfig{file=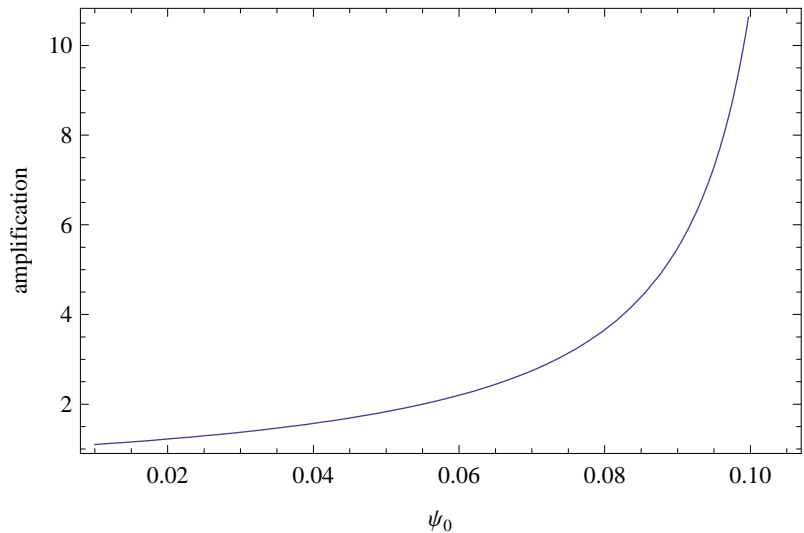}
    \caption{On-axis amplification for the same lensing configuration as for Figure~\ref{F:plummer-v-point-mass-on-axis}, except with the globular cluster M22 modelled as a non-singular isothermal sphere (NSIS) lens, as a function of the characteristic potential $\psi_0$. The amplification rises to a very large (finite) value as $\psi_0 \to \sim 0.11$.}
    \label{F:nsis-on-axis}
  \end{center}
\end{figure}

Figure~\ref{F:nsis-on-axis} is the analogue of Figure~\ref{F:plummer-v-point-mass-on-axis} for the NSIS lens; it shows the on-axis amplification for a NSIS lens at the location of M22 (lensing waves from a source twice as distant as M22), as a function of the parameter $\psi_0$.
For M22 modelled as a NSIS profile, $\psi_0 \simeq 4.1 \times 10^{-16}$, for which the amplification is essentially unity, just as for the Plummer lens. This result is unchanged for NSIS models corresponding to other globular clusters; if globular clusters are well-modelled by NSIS profiles, no significant amplification of LIGO-band gravitational waves is possible.

\section{Effect on detection}
\label{S:detection}

The application of reasonable globular cluster parameters to the various lens models discussed in Section~\ref{S:lens models} shows that, for the SIS lens (and for the unrealistic point-mass lens), significant amplification of LIGO-band gravitational waves is possible under ideal alignment of source, lens, and observer, but for the 2-parameter Plummer and NSIS profiles, which have flat (finite) core densities, no significant amplification is possible under any alignment. We expect most globular clusters to be better modelled as NSIS or Plummer profiles than as SIS profiles. Even if globular clusters are well-modelled as SIS profiles, which may be the case for core-collapsed clusters with a power-law central density, the following argument shows that there is, in any case, a negligible probability of significant lensing, by clusters modelled as SIS profiles, of gravitational waves reaching Earth from an asymmetric neutron star in our galaxy.

Of the 150 known Milky Way globular clusters, only a few lie nearby ($5\,\text{kpc}$) to the Earth and close ($0.5\,\text{kpc}$) to the galactic plane.\footnote{5 globular clusters with structural parameters listed in Harris:\cite{harris_catalog_2003} NGC6540, NGC6544, Terzan12, M22, and M71. Of these, NGC6540 and NGC6544 have collapsed cores.} For M22, very optimistically assuming a density of neutron star sources of $10^6\,\text{kpc}^{-3}$ out to a distance of $15\,\text{kpc}$ behind the lens, and assuming an angular cross section for significant lensing corresponding to the width shown in Figure~\ref{F:sis-M22-diffraction-pattern} ($\sim 2 \times 10^{-7}\,\text{rad}$), we find an upper bound on the number of significantly lensed sources of
\begin{equation}
  \label{E:event-rate}
  \frac{10^6}{\text{kpc}^3} \int_{0\,\text{kpc}}^{15\,\text{kpc}} (2 \times 10^{-7})^2 \pi r^2 \, dr \simeq 1.4 \times 10^{-4},
\end{equation}
where the volume integral is over a cone whose apex is at M22, whose axis of symmetry coincides with the line of sight, and whose half-aperture is $2 \times 10^{-7}\,\text{rad}$. For a typical source velocity of $200\,\text{km}\,\text{s}^{-1}$, the crossing time for the width of Figure~\ref{F:sis-M22-diffraction-pattern} is several years, so Equation~\eqref{E:event-rate} is a rough upper bound on the expected probability of any occurrence of lensing by M22 during a search (using LIGO data, for example) over that time-scale. Although there are several other candidate globular clusters with somewhat different parameters, this cannot account for 4 orders of magnitude, and we conclude that no detectable occurrences of lensing of gravitational waves from asymmetric neutron stars by globular clusters can be expected.

\appendix{Density and surface density of the NSIS profile}
\label{A:nsis-table}

Table~\ref{T:nsis-values} is a more highly accurate version of Table~4-1 of Binney and Tremaine\cite{binney_galactic_1987}, giving selected values of the density and surface density of the NSIS profile as a function of the dimensionless radius $r/r_0$.

\begin{table}
  \tbl{Selected values of the density ($\log_{10} (\rho / \rho_0)$) and surface density ($\log_{10} (\Sigma / r_0 \rho_0)$) for the non-singular isothermal sphere (NSIS) profile, as a function of dimensionless radius $r/r_0$. This is a more highly accurate version of Table~4-1 of Binney and Tremaine\cite{binney_galactic_1987}.}
  {
    \begin{tabular}{| r l l | r l l |}
      \hline
        $r / r_0$ & $\log_{10} (\rho / \rho_0)$ & $\log_{10} (\Sigma / r_0 \rho_0)$ & $r / r_0$ & $\log_{10} (\rho / \rho_0)$ & $\log_{10} (\Sigma / r_0 \rho_0)$ \\
        \hline
        0 & 0 & 0.3050 &  10 & -2.7291 & -1.2137 \\
        0.1 & -0.0065 & 0.3007 & 20 & -3.3217 & -1.4902 \\
        0.2 & -0.0256 & 0.2881 & 30 & -3.6489 & -1.6466 \\
        0.3 & -0.0564 & 0.2677 & 50 & -4.0592 & -1.8486 \\
        0.5 & -0.1468 & 0.2082 & 70 & -4.3345 & -1.9872 \\
        0.7 & -0.2639 & 0.1320 & 100 & -4.6336 & -2.1393 \\
        1 & -0.4618 & 0.0055 &  200 & -5.2347 & -2.4460 \\
        2 & -1.0715 & -0.3645 & 300 & -5.5939 & -2.6282 \\
        3 & -1.5077 & -0.6089 & 500 & -6.0479 & -2.8566 \\
        5 & -2.0560 & -0.8923 & 700 & -6.3456 & -3.0053 \\
        7 & -2.3946 & -1.0565 & 1000 & -6.6590 & -3.1613 \\
       \hline
    \end{tabular}
  }
  \label{T:nsis-values}
\end{table}


\begin{thebibliography}{10}

\bibitem{nakamura_wave_1999}
T.~T. Nakamura and S.~Deguchi, {\em Progress of Theoretical Physics Supplement}
  {\bf 133}, 137 (1999).

\bibitem{nakamura_gravitational_1998}
T.~T. Nakamura, {\em Physical Review Letters} {\bf 80}, 1138 (1998).

\bibitem{ruffa_gravitational_1999}
A.~A. Ruffa, {\em Astrophysical Journal} {\bf 517}, L31 (1999).

\bibitem{baraldo_gravitationally_1999}
C.~Baraldo, A.~Hosoya and T.~T. Nakamura, {\em Physical Review D} {\bf
  59}, 083001 (1999).

\bibitem{takahashi_wave_2003}
R.~Takahashi and T.~Nakamura, {\em Astrophysical Journal} {\bf 595},
  1039 (2003).

\bibitem{takahashi_quasi-geometrical_2004}
R.~Takahashi, {\em Astronomy \& Astrophysics} {\bf 423}, 787 (2004).

\bibitem{macquart_scattering_2004}
J.~P. Macquart, {\em Astronomy \& Astrophysics} {\bf 422}, 761 (2004).

\bibitem{takahashi_amplitude_2006}
R.~Takahashi, {\em Astrophysical Journal} {\bf 644}, 80 (2006).

\bibitem{schneider_gravitational_1992}
P.~Schneider, J.~Ehlers and E.~E. Falco, {\em Gravitational Lenses} (Springer,
  1992).

\bibitem{arnaud-varvella_increase_2004}
M.~Arnaud-Varvella, M.~Angonin and P.~Tourrenc, {\em General Relativity and
  Gravitation} {\bf 36}, 983 (2004).

\bibitem{seto_strong_2004}
N.~Seto, {\em Physical Review D} {\bf 69}, 022002 (2004).

\bibitem{suyama_wave_2005}
T.~Suyama, R.~Takahashi and S.~Michikoshi, {\em Physical Review D} {\bf
  72}, 043001 (2005).

\bibitem{iserles_highly_2005}
A.~Iserles, S.~P. N{{\O}}rsett and S.~Olver, Highly oscillatory quadrature: The
  story so far, in {\em ENuMath, Santiago de Compostela\/},  eds. A.~B.
  de~Castro et~al., {\em Proceedings of ENuMath}, 97 (Springer-Verlag, 2006).

\bibitem{moylan_highly_2007}
A.~J. Moylan, G.~V. Bicknell, D.~E. McClelland, S.~M. Scott and A.~C. Searle
  (2007), in preparation.

\bibitem{levin_procedures_1982}
D.~Levin, {\em Mathematics of Computation} {\bf 38}, 531 (1982).

\bibitem{takahashi_wave_2004}
R.~Takahashi, Wave effects in the gravitational lensing of gravitational waves
  from chirping binaries, PhD thesis, Kyoto University (2004).

\bibitem{ulmer_femtolensing_1995}
A.~Ulmer and J.~Goodman, {\em Astrophysical Journal} {\bf 442}, 67 (1995).

\bibitem{levin_fast_1996}
D.~Levin, {\em Journal of Computational and Applied Mathematics} {\bf 67},
  95 (1996).

\bibitem{levin_analysis_1997}
D.~Levin, {\em Journal of Computational and Applied Mathematics} {\bf 78},
  131 (1997).

\bibitem{olver_moment-free_2006}
S.~Olver, {\em IMA Journal of Numerical Analysis} {\bf 26}, 213 (2006).

\bibitem{olver_quadrature_2006}
S.~Olver, {\em Numerische Mathematik} {\bf 103}, 643 (2006).

\bibitem{chung_method_2000}
K.~C. Chung, G.~A. Evans and J.~R. Webster, {\em Applied Numerical Mathematics}
  {\bf 34}, 85 (2000).

\bibitem{harris_catalog_2003}
W.~E. Harris, Catalog of milky way globular clusters, \\ http://physwww.mcmaster.ca/\%7Eharris/Databases.html (2003).

\bibitem{pryor_velocity_1993}
C.~Pryor and G.~Meylan, Velocity dispersions for galactic globular clusters, in
  {\em Structure and Dynamics of Globular Clusters\/},  eds. S.~G. Djorgovski
  and G.~Meylan, {\em ASP Conference Series} {\bf 50}, 357 (1993).

\bibitem{peterson_proper_1994}
R.~C. Peterson and K.~M. Cudworth, {\em Astrophysical Journal} {\bf 420}, 612
  (1994).

\bibitem{peters_index_1974}
P.~C. Peters, {\em Physical Review D} {\bf 9}, 2207 (1974).

\bibitem{plummer_problem_1911}
H.~C. Plummer, {\em Monthly Notices of the Royal Astronomical Society} {\bf
  71}, 460 (1911).

\bibitem{binney_galactic_1987}
J.~Binney and S.~Tremaine, {\em Galactic Dynamics} (Princeton University Press,
  1987).

\end{thebibliography}

\end{document}